\documentclass{ws-ijmpa}

\begin{document}

\markboth{TIM HUEGE}{SIMULATIONS OF RADIO EMISSION FROM CR AIR SHOWERS}

\catchline{}{}{}{}{}

\title{SIMULATIONS OF RADIO EMISSION FROM COSMIC RAY AIR SHOWERS}

\author{TIM HUEGE}
\address{Max-Planck-Institut f\"ur Radioastronomie, Auf dem H\"ugel 69, 53121 Bonn, Germany}

\author{HEINO FALCKE}
\address{Radio Observatory, ASTRON, Dwingeloo, PO Box 2, 7990 AA Dwingeloo, The Netherlands}  


\maketitle

\pub{Received (Day Month Year)}{Revised (Day Month Year)}

\begin{abstract}
Radio emission from cosmic ray air showers has the potential to become an additional, cost-effective observing technique for cosmic ray research, being largely complementary to the well-established particle detector and air fluorescence techniques. We present Monte Carlo simulatios of radio emission from extensive air showers in the scheme of coherent geosynchrotron radiation from electron-positron pairs gyrating in the earth's magnetic field. Preliminary results of our simulations are the predicted frequency, primary particle energy, shower zenith angle, shower azimuth angle and polarization dependence of the radio emission. These properties can be directly related to data measured by LOPES and other experiments.
\end{abstract}



\section{Introduction}

Accompanying the experimental efforts of the LOPES project\cite{Horneffer2004}, we are modeling the radio emission from cosmic ray air showers in the scheme of coherent geosynchrotron radiation\cite{FalckeGorham2003}, thereby building a theoretical foundation for the interpretation of experimental data. As a follow-up to our analytic analysis of the emission properties and the underlying coherence effects\cite{HuegeFalcke2003a}, we have carried out sophisticated Monte Carlo simulations based on analytic parametrizations of the air shower properties to achieve a high-precision modeling of the emission including its polarization characteristics\cite{HuegeFalcke2004a}. Here, we present some of our preliminary results.

\section{Monte Carlo code}

Our Monte Carlo code calculates the geosynchrotron radiation from the air shower cascade's electron-positron pairs gyrating in the earth's magnetic field. All calculations are done in the time-domain, without the application of any far-field approximations. The full polarization information of the emission is preserved. This approach is robust, fast and flexible and has been thoroughly tested\cite{HuegeFalcke2004a}. The underlying air shower model is based on analytic parametrizations\cite{HuegeFalcke2003a} and takes into account longitudinal and lateral particle distributions, particle track length and energy distributions, the air shower and magnetic field geometry and the air shower evolution as a whole. This allows a direct comparison of the analytic and Monte Carlo results. At a later time, the air shower model will be substituted by a direct interfacing to the air shower simulation code CORSIKA. Additional emission mechanisms (e.g., Askaryan-type \v{C}erenkov-radiation) can be included at a later time.

\section{Simulation Results}

In this section, we present a number of preliminary results of our simulations.

\subsection{Frequency Spectra}

The spectral dependence of the emission (not shown here) clearly favors low frequencies for the measurement of radio emission from EAS. At a frequency of 10~MHz, the emission stays coherent up to high distances ($\sim$~1,000~m) from the shower centre. Additionally, the signal strength is higher than at higher frequencies. At a frequency of 55~MHz, coherence holds up to $\sim300$~m for a vertical air shower. Observing EAS at larger distances, i.e.\ in the incoherent regime, is still likely to be possible, but Monte Carlo simulations with a more sophisticated air shower model as will be achieved through interfacing with CORSIKA are needed.

\subsection{Primary particle energy}

In the coherent frequency regime, e.g.\ at 10~MHz, the electric field strength scales almost linearly with the primary particle energy (not shown here). This is the expected behavior for coherent emission. If an increase in depth of shower maximum as a function of primary particle energy is taken into account, the energy-scaling becomes dependent on the distance from the shower maximum. While it stays a power-law at all times, the power-law index drops from $\sim 1.1$ in the centre region to $\sim 0.8$ at 500~m distance.

\subsection{Polarization characteristics}

\begin{figure}
\begin{center}
a)\hspace{-.3cm}\begin{minipage}[t]{0.42\linewidth}
\vspace{.05\linewidth}\psfig{file=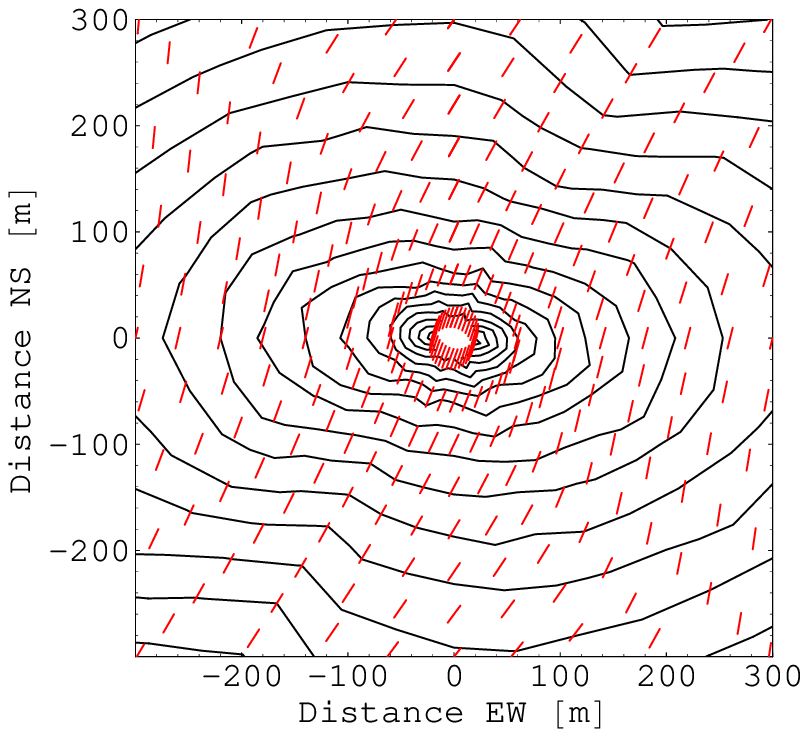,width=\linewidth}
\end{minipage}
\hspace{0.2cm}
b)\hspace{-0.4cm}\begin{minipage}[t]{0.54\linewidth}
\vspace{.05\linewidth}\psfig{file=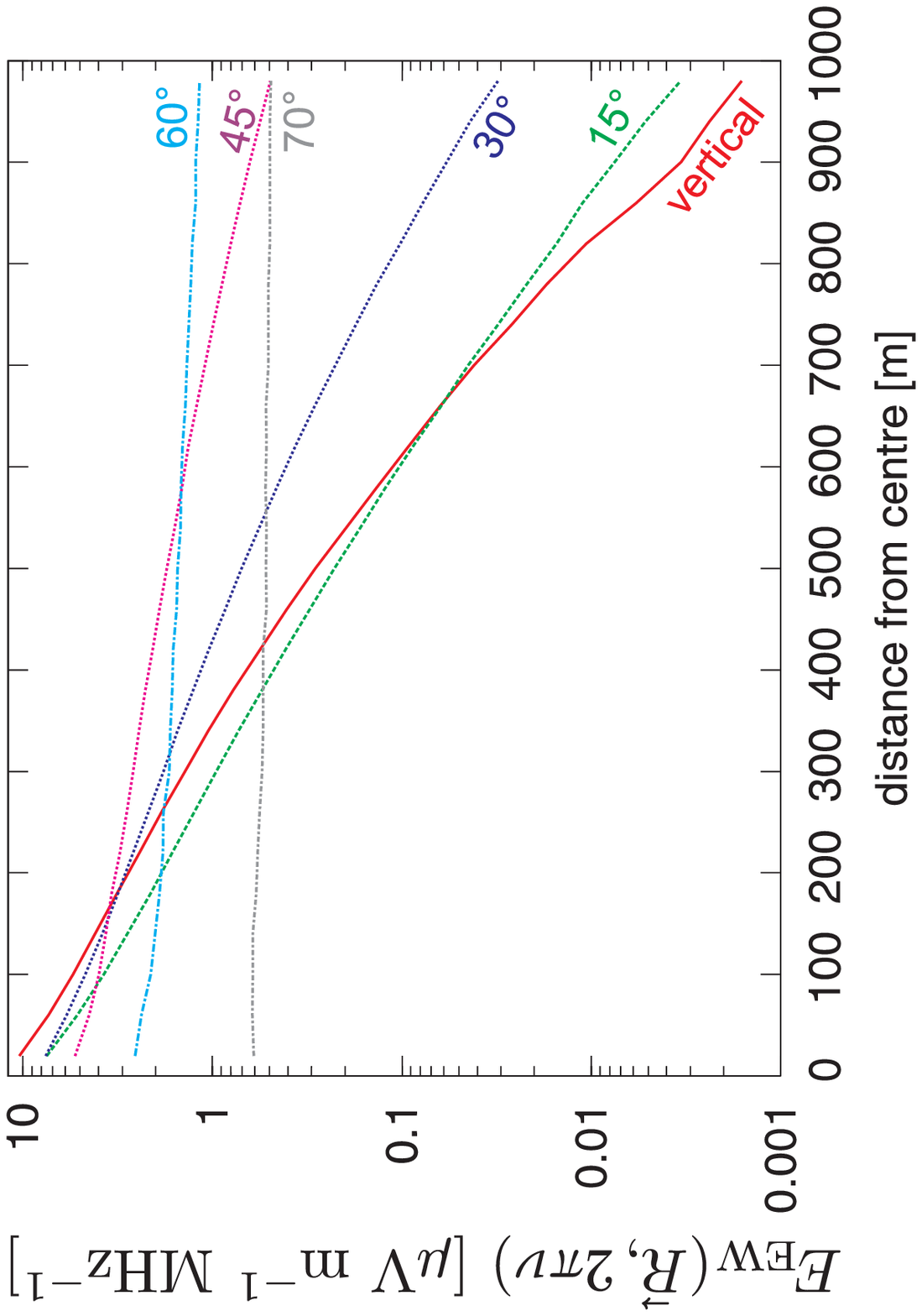,angle=-90,width=\linewidth}
\end{minipage}
\caption{
 \label{figure}
a) Total 10 MHz field strength contours for a $10^{17}$~eV air shower with 45$^{\circ}$ zenith angle coming from the east, overlaid with polarization indicators denoting the ratio between east-west and north-south linear polarization. b) Zenith angle dependence of the 10~MHz emission.
}
\end{center}
\end{figure}

Figure \ref{figure}a) shows the total field strength contours for a 45$^{\circ}$ zenith angle air shower. The emission pattern is elliptic due to projection effects arising from the inclination of the shower axis. Overlaid are indicators denoting the ratio between east-west and north-south linear polarization. As the total field strength pattern is simply aligned with the air shower axis and does not reveal any significant dependence on the geomagnetic field direction, it cannot be used for a verification of the geomagnetic emission mechanism. However, the emission is highly polarized in the direction perpendicular to the shower and geomagnetic field axes. Experiments measuring linear polarization can thus directly test the hypothesis of geomagnetic emission.

\subsection{Zenith angle}

As can be seen in Fig.\ \ref{figure}b), the radial dependence of the emission flattens significantly with increasing zenith angle. Apart from simple projection effects, an intrinsic flattening occurs as a consequence of the growing distance of the shower maximum for increasing zenith angles. The slight deviations from the trend visible for the 15$^{\circ}$ case (i.e., at only 5$^{\circ}$ relative to the 70$^{\circ}$ inclined geomagnetic field) demonstrates the weak influence of the geomagnetic field on the field strength.

\section{Conclusions}

We have developed a powerful Monte Carlo code for the simulation of radio emission from cosmic ray air showers. Our predicted dependence of the radio emission on important air shower parameters provides a necessary prerequisite for cosmic ray research with radio techniques. In particular, the predicted polarization characteristics allow a direct verification of the geomagnetic origin of the radiation with LOPES. A detailed analysis of our results is in preparation.

\section*{Acknowledgements}

Tim Huege was supported by the International
Max Planck Research School (IMPRS) for Radio and Infrared Astronomy at the
University of Bonn.


\end{document}